\documentclass[fleqn,usenatbib]{mnras}

\usepackage{newtxtext,newtxmath}
\usepackage{xcolor}
\usepackage{ulem} 
\usepackage[T1]{fontenc}
\DeclareRobustCommand{\VAN}[3]{#2}
\let\VANthebibliography\thebibliography
\def\thebibliography{\DeclareRobustCommand{\VAN}[3]{##3}\VANthebibliography}

\usepackage{graphicx}	
\usepackage{amsmath}	

\title[Single pulses of PSR J2222$-$0137 with FAST]{Variability, polarimetry, and timing properties of single pulses from PSR~J2222$-$0137 using FAST}   

\author[X. L. Miao et al.]{X. L. Miao$^{1}$\href{https://orcid.org/0000-0003-1185-8937}{\includegraphics[scale=0.08]{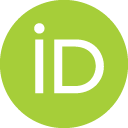}},\thanks{E-mail: xlmiao@nao.cas.cn}
W. W. Zhu$^{1,2}$\href{https://orcid.org/0000-0001-5105-4058}{\includegraphics[scale=0.08]{ORCIDiD.png}},\thanks{E-mail: zhuww@nao.cas.cn}
M.~Kramer$^{3,4}$\href{https://orcid.org/0000-0002-4175-2271}{\includegraphics[scale=0.08]{ORCIDiD.png}}, 
P.~C.~C.~Freire$^{3}$\href{https://orcid.org/0000-0003-1307-9435}{\includegraphics[scale=0.08]{ORCIDiD.png}}, 
L. Shao,$^{5,1,3}$\href{https://orcid.org/0000-0002-1334-8853}{\includegraphics[scale=0.08]{ORCIDiD.png}}, 
M.~Yuan$^{1,7}$\href{https://orcid.org/0000-0003-1874-0800}{\includegraphics[scale=0.08]{ORCIDiD.png}},
\newauthor
L.~Q.~Meng$^{1,7}$\href{https://orcid.org/0000-0002-2885-568X}{\includegraphics[scale=0.08]{ORCIDiD.png}}, 
Z.~W.~Wu$^{1}$\href{https://orcid.org/0000-0002-1381-7859}{\includegraphics[scale=0.08]{ORCIDiD.png}},
C.~C.~Miao$^{1,7}$\href{https://orcid.org/0000-0002-9441-2190}{\includegraphics[scale=0.08]{ORCIDiD.png}}, 
Y. J. Guo$^{3}$,
D.~J.~Champion$^{3}$\href{https://orcid.org/0000-0003-1361-7723}{\includegraphics[scale=0.08]{ORCIDiD.png}}, 
E.~Fonseca$^{8,9,10,11}$\href{https://orcid.org/0000-0001-8384-5049}{\includegraphics[scale=0.08]{ORCIDiD.png}},
\newauthor
J.~M.~Yao$^{12}$,
M.~Y.~Xue$^{1}$\href{https://orcid.org/0000-0001-8018-1830}{\includegraphics[scale=0.08]{ORCIDiD.png}},
J.~R.~Niu$^{1,7}$\href{https://orcid.org/0000-0001-8065-4191}{\includegraphics[scale=0.08]{ORCIDiD.png}},
H. Hu$^{3}$\href{https://orcid.org/0000-0002-3407-8071}{\includegraphics[scale=0.08]{ORCIDiD.png}},
C.~M.~Zhang$^{1}$
\\
$^{1}$National Astronomical Observatories, Chinese Academy of Sciences, 20A Datun Road, Chaoyang District, Beijing 100101, China \\
$^{2}$Institute for Frontiers in Astronomy and Astrophysics, Beijing Normal University, Beijing 102206, China\\
$^{3}$Max-Planck Institut f{\"u}r Radioastronomie, Auf dem H{\"u}gel 69, D-53121 Bonn, Germany\\ 
$^{4}$Jodrell Bank Centre for Astrophysics, Department of Physics and Astronomy, University of Manchester, M13 9PL Manchester, UK\\
$^{5}$Kavli Institute for Astronomy and Astrophysics, Peking University, Beijing 100871, China \\
$^{6}$Department of Astronomy, School of Physics, Peking University, Beijing 100871, China \\
$^{7}$School of Astronomy and Space Science, University of Chinese Academy of Sciences, Beijing, 100049, China \\ 
$^{8}$Department of Physics and Astronomy, West Virginia University, PO Box 6315, Morgantown, WV 26506, USA\\
$^{9}$Center for Gravitational Waves and Cosmology, West Virginia University, Chestnut Ridge Research Building, Morgantown, WV 26505, USA\\
$^{10}$Department of Physics, McGill University, 3600 rue University, Montr{\'e}al, QC H3A 2T8, Canada\\
$^{11}$McGill Space Institute, McGill University, 3550 rue University, Montr{\'e}al, QC H3A 2A7, Canada\\
$^{12}$Xinjiang Astronomical Observatory, Chinese Academy of Sciences, Urumqi, Xinjiang 830011, People's Republic of China \\ 
}

\date{Accepted XXX. Received YYY; in original form ZZZ}

\pubyear{2023}

\begin{document}
\label{firstpage}
\pagerange{\pageref{firstpage}--\pageref{lastpage}}
\maketitle

\begin{abstract}
In our work, we analyse $5\times10^{4}$ single pulses from the recycled pulsar PSR~J2222$-$0137 in one of its scintillation maxima observed by the Five-hundred-meter Aperture Spherical radio Telescope (FAST).
PSR~J2222$-$0137 is one of the nearest and best studies of binary pulsars and a unique laboratory for testing gravitational theories.
We report single pulses' energy distribution and polarization from the pulsar's main-pulse region.
The single pulse energy follows the log-normal distribution.
We resolve a steep polarization swing, but at the current time resolution ($64\,\mu{\rm s}$), we find no evidence for the orthogonal jump in the main-pulse region, as has been suspected.
We find a potential sub-pulse drifting period of {$P_{3} \sim 3.5\,P$}.
We analyse the jitter noise from different integrated numbers of pulses and find that its $\sigma_{j}$ is {$270\pm{9}\,{\rm ns}$ }for 1-hr integration at 1.25 GHz. This result is useful for optimizing future timing campaigns with FAST or other radio telescopes.
\end{abstract}




\section{Introduction}

\begin{figure*}
    \includegraphics[width=16cm]{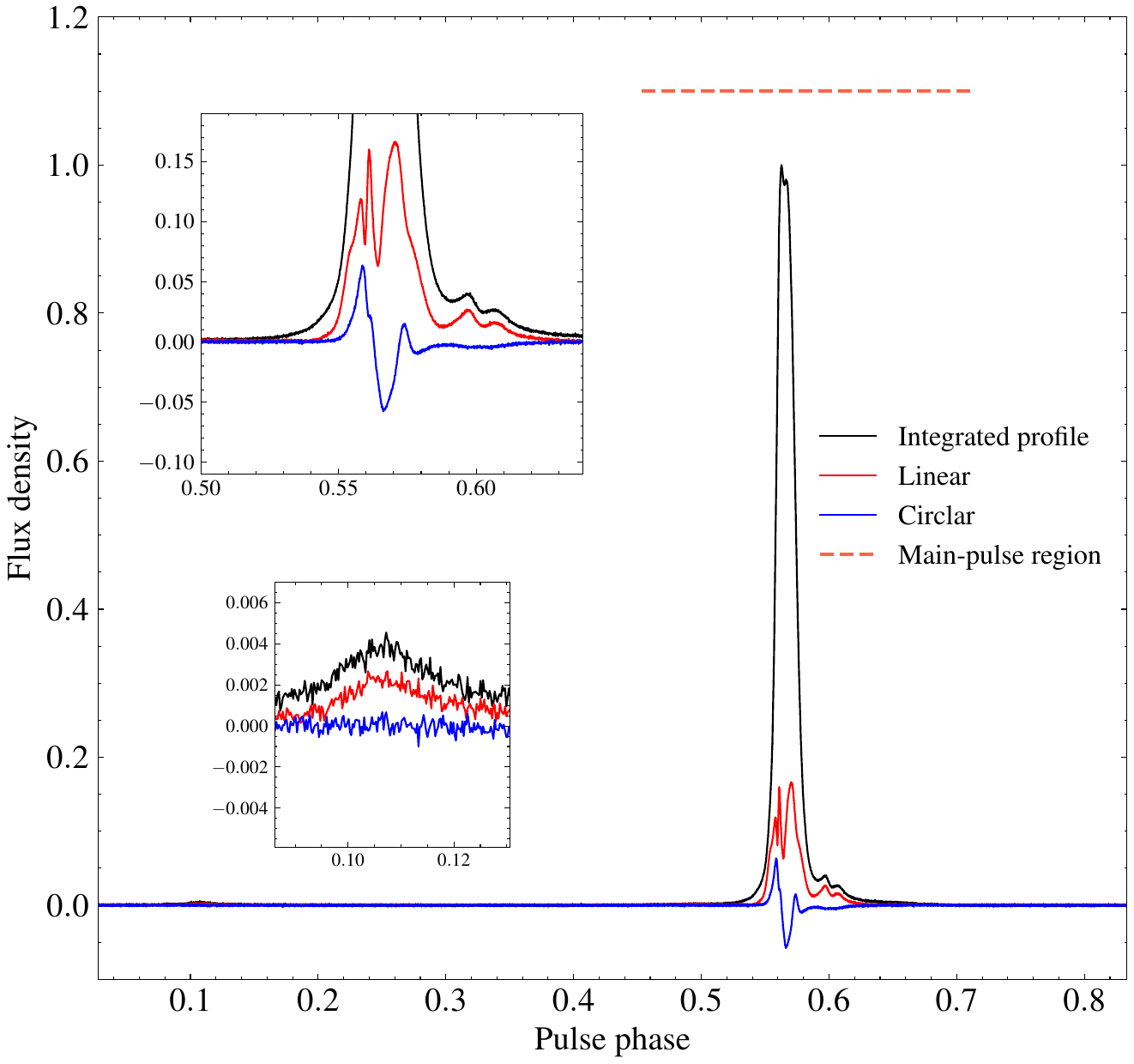}
    \centering
    \caption{The polarization profile of PSR~J2222$-$0137 is observed by FAST at a central frequency of 1250 MHz on MJD 59209, and it is averaged over a bandwidth of 420 MHz. The black line is the total intensity (SNR$\sim$23000) of the integration time of 30 min, the red line is the linear polarization intensity, and the blue line is the circular polarization intensity. The upper inset shows an enlarged version of the main pulse, and the bottom inset shows an enlarged version of the interpulse which was first detected by FAST.
    The red dashed line marks the selected main-pulse region used for single-pulse analysis.
    \label{fig:integrationprofile}}
\end{figure*}

Millisecond pulsars (MSPs, defined here in the broad sense of having been recycled by accretion of matter from their companions, see \citealt{2023pbse.book.....T}) are neutron stars (NSs) known for their superb rotational stability.
Like most pulsars, their integrated profiles are also stable; this means that we can obtain precise pulses' times of arrival {\citep[TOAs; there are also some exceptions, such as PSRs J1643$-$1224, J1713+0747, and see
e.g.][]{Shannon_2016ApJ,Xu_2021ATel,Singha_2021MNRAS}}. 
This, combined with the fact that most MSPs are in binary systems, implies that we can use TOA analysis (pulsar timing) to measure their orbital motions with exquisite precision.

In some binary pulsars, consistent pulsar timing can detect - and precisely measure - relativistic perturbations to the orbital motion and the propagation of radio waves in the system. 
If enough relativistic effects are detected in the same system, we can obtain rigorous tests of gravity theories (\citealt{1982ApJ...253..908T,1989ApJ...345..434T,1992PhRvD..45.1840D,2012MNRAS.423.3328F,2020A&A...638A..24V,2021PhRvX_kramer,Shao_2022arXiv}, for reviews see \citealt{Wex:2014nva,Hu_2023arXiv}). 
For a broader class of binaries, the measurement of these relativistic effects allows measurements of NS masses, which is important for several reasons, mostly the study of the unknown properties of the super-dense matter at the centre of a NS \citep{2016ARA&A..54..401O}. 
Irrespective of the detection of relativistic effects in their orbits, bright MSPs are also crucial for nHz-{gravitational wave (GW)} detection by pulsar timing arrays (PTAs, \citealt{Estabrook_1975GReGr,Detweiler:1979wn,Hellings_1983ApJ,Haasteren_2009MNRAS}); in these experiments, the TOA accuracy determines {the sensitivity of PTAs to nHz-GWs detection. Recently, four independent PTA groups have found the critical evidence of nHz-GW background \citep{Xu_2023RAA,Reardon_2023ApJ,Antoniadis_2023arXiv,Agazie_2023arXiv}, and we are in the era of opening the nHz-GW observation window.}

\subsection{PSR J2222$-$0137: a nearby laboratory for gravity theories}

Among the best laboratories for tests of gravity theories is PSR~J2222$-$0137, a recycled pulsar with a massive white dwarf (WD) companion in a 2.44-day orbit \citep{2014ApJ_Kaplan}.
The system has an edge-on orbit, allowing a highly precise measurement of the Shapiro delay \citep{1964PhRvL..13..789S}, which provides an accurate mass measurement for the pulsar and its WD companion,
$1.831 \pm 0.010 \, \rm M_{\odot}$ and $1.319 \pm 0.004 \, \rm M_{\odot}$ respectively. 
The large masses for the pulsar and WD companion make it the most massive double degenerate binary pulsar known in our Galaxy, with a total mass of $3.150 \pm 0.014 \, \rm M_{\odot}$. In this system, we can also obtain a significant measurement of the rate of advance of periastron $\dot{\omega}$, which provides a redundant measurement of the system's total mass and a 1\% test of GR, which the theory passes \citep{Guo_2021}.

Furthermore, being one of the nearest binary pulsars, it has a precise distance measurement, $d=268.1^{+1.2}_{-1.1}\,{\rm pc}$, by Very Long Baseline Interferometry \citep{2013ApJ_Deller,Guo_2021}. This and the precise timing results in a precisely measured intrinsic orbital period variation, {$\dot{P}_{b}^{\rm intr}=-0.0143\pm{0.0076}\times10^{-12}\,{\rm s\,s^{-1}}$ }\citep{Guo_2021}.

These measurements are important because the close agreement between $\dot{P}_{b}^{\rm intr}$ and the GR prediction provides a tight limit on the emission of dipolar gravitational waves (DGW) from this system \citep{Guo_2021,batrakov2023new}, a phenomenon generally predicted by alternative theories of gravity \citep{1975ApJ...196L..59E}.
Furthermore, the pulsar mass places it in a NS mass range that had previously not been probed by precise tests of GW emission, where the phenomenon of spontaneous scalarisation (a large increase of DGW emission predicted by some alternative gravity theories, like those of \citealt{Damour_1993PhRv}) was still possible. For this reason, the DGW emission limit for PSR~J2222$-$0137 represents a powerful constraint on that phenomenon \citep{2022CQGra_zhao,batrakov2023new}.

\subsection{FAST observations of PSR J2222$-$0137}

The high sensitivity of the Five-hundred-meter Aperture Spherical radio Telescope (FAST) allows us to measure the emission of PSR~J2222$-$0137 with high precision. The main aim of these observations is to improve the precision of the tests of gravity theories with this pulsar.
However, even before that, the FAST data from this pulsar - particularly during a scintillation maximum - have already provided an extremely high signal-to-noise ratio (SNR) integrated profile. This resulted in an improved polarization study and revealed the existence of an interpulse for the first time. These measurements were crucial for determining the large-scale structure of the magnetic field of the pulsar, which allowed an estimate of the spin and orbital geometry of this pulsar - which was confirmed independently by timing measurements \citep{Guo_2021}.

The scintillation maxima data from PSR~J2222$-$0137 is also useful for a single-pulse analysis.
In general, the integration of hundreds or thousands of pulse periods leads to stable profiles; this is the fundamental assumption of pulsar timing, which allows us to measure pulsar parameters with high precision. 
However, the single-pulse shape changes stochastically from pulse to pulse, and some of them exhibit special temporal characteristics, like sub-pulses drifting \citep{1968Natur_Drake,2006A_A_Weltevrede}, mode change \citep{1982ApJ_Bartel}, nulling \citep{1970Natur_backer,2003A_A_Edwards,2007MNRAS_Wang}, giant pulses \citep{1968Sci_stealin,2003Nature_hankins}, and
microstructure in single pulses \citep{1990AJ_Cordes,Mitra_2015ApJ,Kishalay_2016ApJ,Liu_2022MNRAS}.
These temporal characteristics in single pulses are often found in normal pulsars but are uncommon in MSPs.

This pulse-to-pulse variability results in the pulse phase jitter phenomenon in integrated profiles \citep{Cordes_1985,Cordes_1993,Cordes_2010}, a trivial consequence of the fact that it takes significant time and pulse averaging for the integrated pulse to get close to ``the'' average pulse profile of the pulsar. 
These small variations of the integrated pulse profile limit the timing precision of MSPs and influence the ability of test gravity and PTAs to detect nHz GW.

{A} comprehensive study of single pulse variability in MSPs {could help understand} the limits of the timing precision \citep{2011MNRAS_Oslowski,2015MNRAS_Imgrund}. 
For a long time, the limited telescope sensitivity of most radio telescopes implied that only the single pulses of a handful of the brightest MSPs could be studied \citep{shannon_2012ApJ,Liu_2014}.
The high sensitivity of FAST allows us to investigate more MSPs, and these investigations can lead to advances not only in understanding the limitations of their timing precision, but also in our understanding of the emission physics of pulsars.
For PSR~J2222$-$0137, the jitter phenomenon is evident in high SNR observations from FAST.
Therefore, it is important to study the jitter noise of PSR~J2222$-$0137 to optimise future timing campaigns with FAST or other high-sensitivity radio telescopes.

As mentioned above, the polarization profiles of pulsars can provide important information about their geometry. A commonly used model is the rotating vector model (RVM, \citealt{1969_Radhakrishnan}), which describes the variations in the linear polarization position angle (PPA) of many pulsars well.
However, some pulsars' linear PPA show rapid jumps of $90^{\circ}$, which deviate from a standard RVM. 
Such jumps are believed to come from the presence of two orthogonal polarization modes in the radiation and are called orthogonal polarization modes (OPMs) \citep{1995MNRAS_Gil,2004A_AEdwards}. 
\citet{Guo_2021}'s figure 1 shows the integrated polarization profile of PSR~J2222$-$0137, and the authors suggest that the marked grey region of the {PPA} of the linear emission results from OPMs.
In this work, we will analyse this particular polarization jump by resolving it through single pulse polarization measurements and investigate the above interpretation.

In this paper, we present a comprehensive analysis using the high SNR data of PSR~J2222$-$0137.
In Section \ref{sec:obs}, we describe the details of the observations and single pulse data analysis. In Section \ref{sec:res}, we show the results of this analysis, including the single pulse energy distribution, pulse variability and polarization, and finally, an analysis of the integrated pulse jitter for PSR~J2222$-$0137. We conclude and discuss our results in Section \ref{sec:con}.

\section{Observation and Data Processing}\label{sec:obs}

\begin{figure}
    \includegraphics[width= \columnwidth]{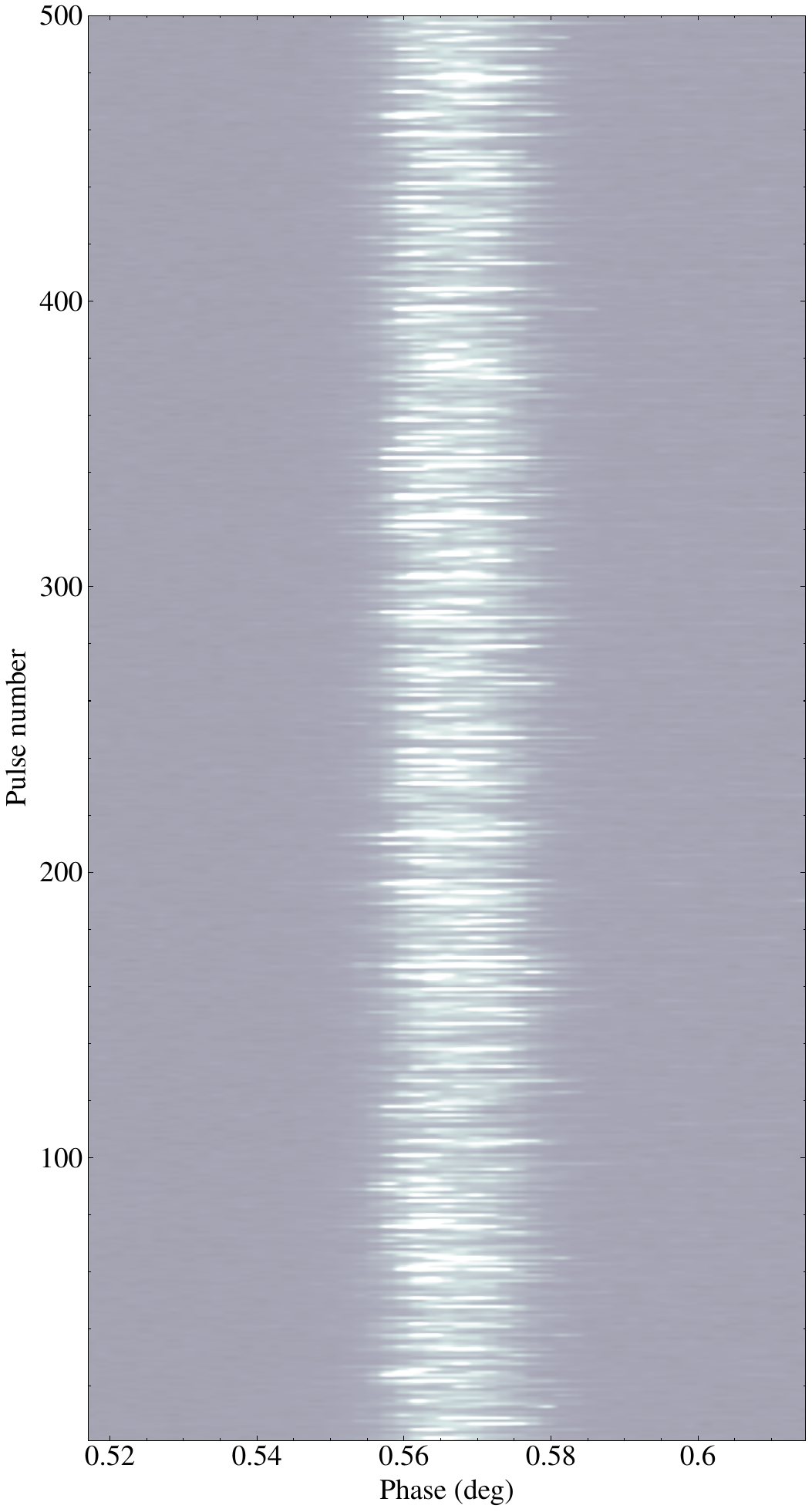}
    \caption{The pulse stack of PSR~J2222$-$0137, we show the pulse-to-pulse variation of 500 continuous pulses, for a total duration of 16.4\,s. The time resolution is $64.06\,\mu{\rm s}$, corresponding to the spin period divided by the 512 bins. \label{fig:stack}}
\end{figure}

In our FAST observation project, we have performed thirty 30-min observations of PSR~J2222$-$0137.
We used the central beam of the 19-beam receiver and the frequency range between 1000 and 1500 MHz with 4096 frequency channels. We used only the band between 1040 to 1460 MHz because of the decrease in sensitivity at the edges of the band \citep{2020RAA_Jiang}.
We recorded the data in pulsar search mode with full-Stokes information, and each observation starts with a 1-min calibration noise diode for the polarization calibration.

The pulsar often exhibits a SNR $\sim4000$ in a regular FAST observation. In this paper, we choose the brightest observation of PSR~J2222$-$0137 from MJD 59209, when the pulsar was in a scintillation maximum with a SNR $\sim23000$. 
\cite{Guo_2021} already used this observation to improve the geometric parameters for the pulsar spin.
Fig.\,\ref{fig:integrationprofile} shows the integrated polarization profile of PSR~J2222$-$0137 obtained from this observation. The interpulse is visible between spin phases $0.09-0.13$ and is more clearly visible in the lower inset.

To get single pulses of PSR~J2222$-$0137, we used the DSPSR\footnote{\url{http://dspsr.sourceforge.net}} software package \citep{2011PASA_van} to extract individual pulses and remove inter-channel dispersion delays with the ``-K'' option.
The data were coherently dedispersed using a dispersion measure (DM) of $3.28\,{\rm cm^{-3}\,pc}$ \citep{Guo_2021}.
The spin period of {PSR} J2222$-$0137 is $32.81\,{\rm ms}$ and the time resolution we used is $49.152\,{\mu \rm s}$, so we set the number of phase bins to 512 for every pulse.
We used the PSRCHIVE\footnote{\url{http://psrchive.sourceforge.net}} software package \citep{2004PASA_Hotan} to excise radio frequency interference (RFI) in every single pulse in the frequency domain. 
{We used ``psrzap'' for every single pulse to excise RFI in the frequency domain. 
We recorded excised frequency channels for all single pulses. 
We combined the number of excised frequency channels
and used the combined frequency channels to excise every single pulse again, ensuring that every pulse's excised frequency channels are the same.}
{We used the 1-min calibration noise files to calibrate polarization.}
We used ``rmfit'' to fit the Faraday rotation measure (RM) and got ${\rm RM} = 4.28\pm0.18\,{\rm rad\,m^{-2}}$. We derotated the stokes parameters according to this RM for every single pulse.

In our work, we concentrate on investigating the main-pulse properties, i.e., the spin phases between $0.45-0.71$ marked by the red dashed line in Fig.\,\ref{fig:integrationprofile}, and $0.72-0.98$ as the off-pulse region to provide the information on the noise. 
We obtained 52692 high-SNR single pulses from the main-pulse component. Fig.\,\ref{fig:stack} shows an example of a pulse stack containing 500 consecutive pulses of the main pulse. 
It clearly shows the variation of intensity, shape, and phase from pulse to pulse. 

We used the PSRSALSA\footnote{\url{https://github.com/weltevrede/psrsalsa}} software \citep{2016A_A_Weltevrede} to analyse these single pulses.
We used it to obtain the distribution of the main pulse and noise energy.
We also used PSRSALSA to analyse the sub-pulse drifting. {We used TEMPO\footnote{\url{https://tempo.sourceforge.net}} software to get pulse TOAs and timing residuals, and we used these timing residuals to do jitter analysis.}

\begin{figure}
    \begin{center}
      \includegraphics[width= \columnwidth]{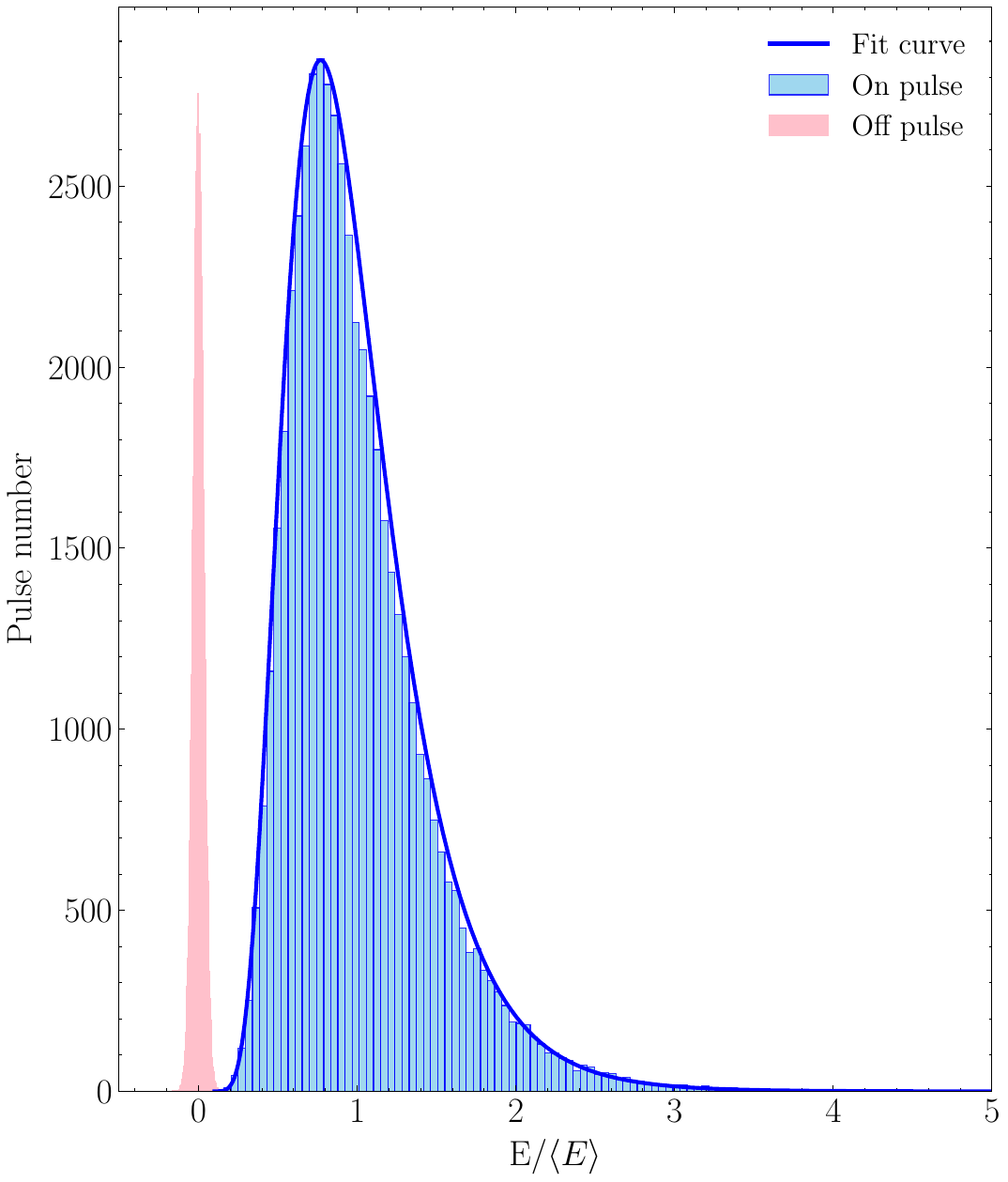}
    \end{center}
\caption{The energy distributions in the main pulse and the off-pulse regions. The sky-blue histogram is the main-pulse energy distribution, and the pink histogram is the noise energy distribution. The blue {solid} line is the fitted curve of the main-pulse energy distribution. The average pulse energy represents, in the observation being analysed, a SNR of 87.}
\label{fig:energy}
\end{figure}

\begin{figure}
    \includegraphics[width= \columnwidth]{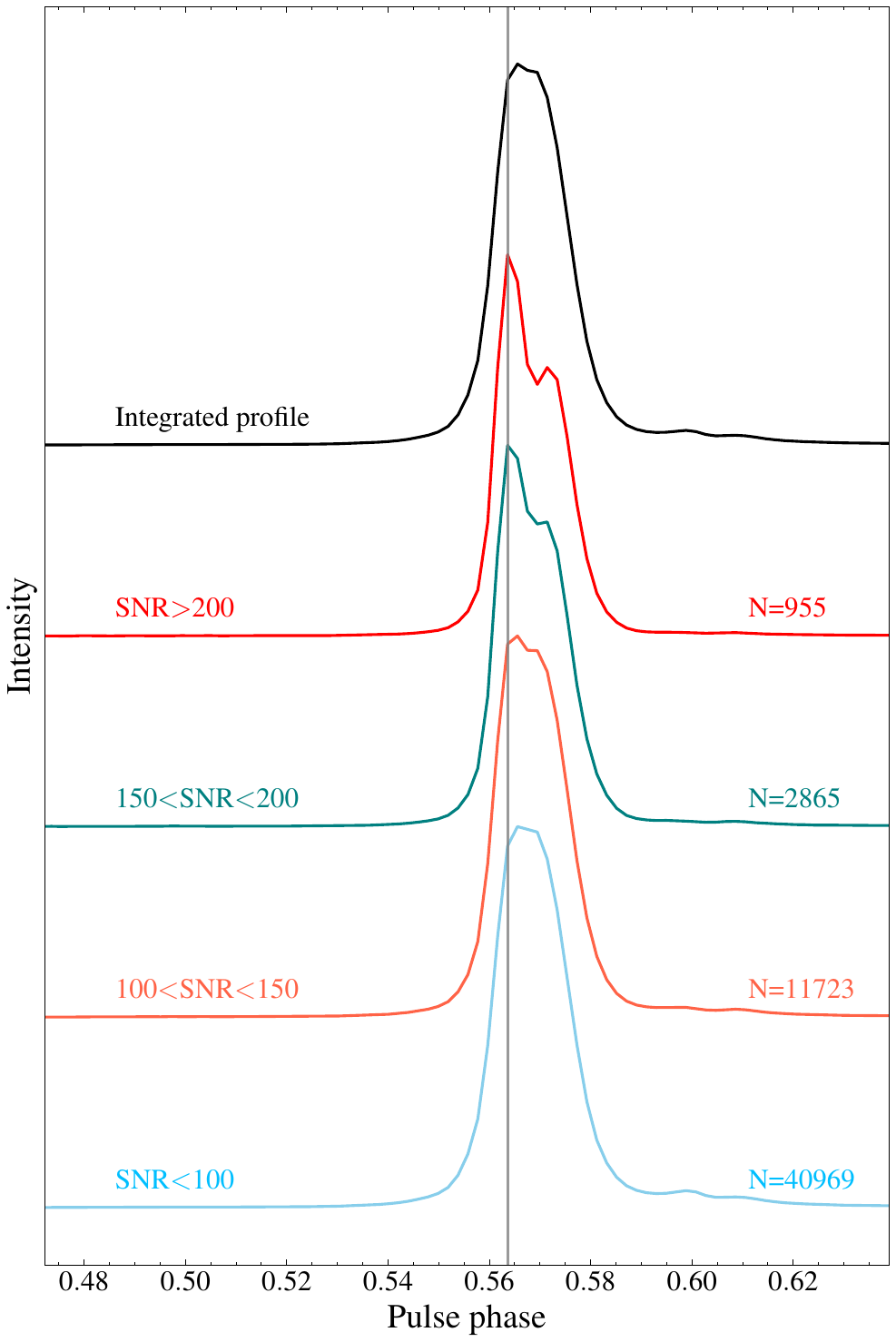}
    \caption{The average pulse profiles for different SNR ranges.
    The profiles are integrated by single pulses with different SNR ranges.
    The black solid line is the total intensity profile, the red solid line is the average pulse profile from single pulses with ${\rm SNR}>200$, the teal solid line is the average pulse profile from single pulses with $150<{\rm SNR}<200$, the tomato solid line is the average pulse profile from single pulses with $100<{\rm SNR}<150$, the sky-blue solid line is the average pulse profile from single pulses with ${\rm SNR}<100$. 
    The grey solid vertical line marks the leading peak of the average pulse profile from single pulses with ${\rm SNR}>200$.
    {The number of pulses averaged N of each SNR range is marked in the figure.}
    The profiles are normalized by the peak value of every average pulse profile individually.
    \label{fig:snrdiffer}}
\end{figure}

\section{Results}\label{sec:res}
This section presents the results of the single pulse analysis of PSR~J2222$-$0137.
In Sec.\,\ref{sec:res:pro}, we provide the energy distributions of these single pulses, which we check for mode change, nulling, sub-pulse drifting, and giant pulse emission.
In Sec.\,\ref{sec:res:polar}, we analyse the single pulse polarization.
In Sec.\,\ref{sec:res:jitter}, we provide a jitter noise investigation of PSR~J2222$-$0137.

\subsection{Single pulse properties}\label{sec:res:pro}

During the MJD 59209 observation, the minimal single pulse SNR is $\sim8.7$, and the maximal is $\sim531.6$.
{These values of SNR are calculated by,}
\begin{equation}\label{eq:snrdefine}
    {\rm SNR} = \frac{\sum I_{i}}{\sigma_{\rm off} \sqrt{n}}\,,
\end{equation}
{where $\sum I_{i}$ ($I_{i}>3\sigma_{\rm off}$) is the sum of the intensity of the on-pulse region of single pulses, $n$ is the number of sum bins, and $\sigma_{\rm off}$ is the Root-Mean-Square (RMS) intensity of the off-pulse region.}
All single pulses show a SNR\,$>8\sigma$, in other words, we do not detect nulling.
\begin{figure}
    \includegraphics[width= \columnwidth]{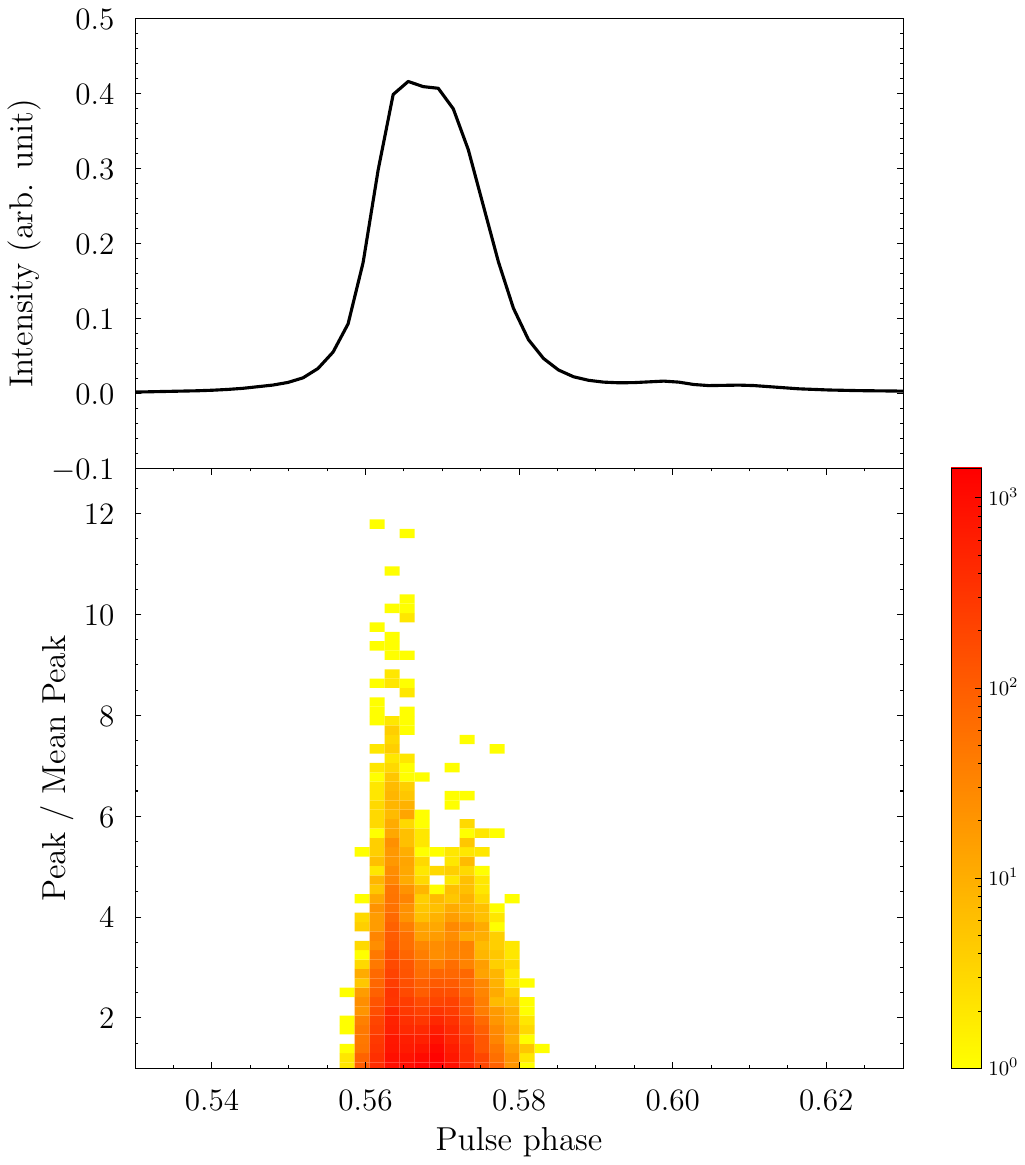}
    \caption{The upper panel is the total intensity profile. The lower panel is the number density distribution of pulses with respect to their peak flux density and the spin phase where that peak occurs. The time of each bin corresponds to $64.06\,{\rm \mu s}$.\label{fig:peakdis}}
\end{figure}
\begin{figure}
    \includegraphics[width= \columnwidth]{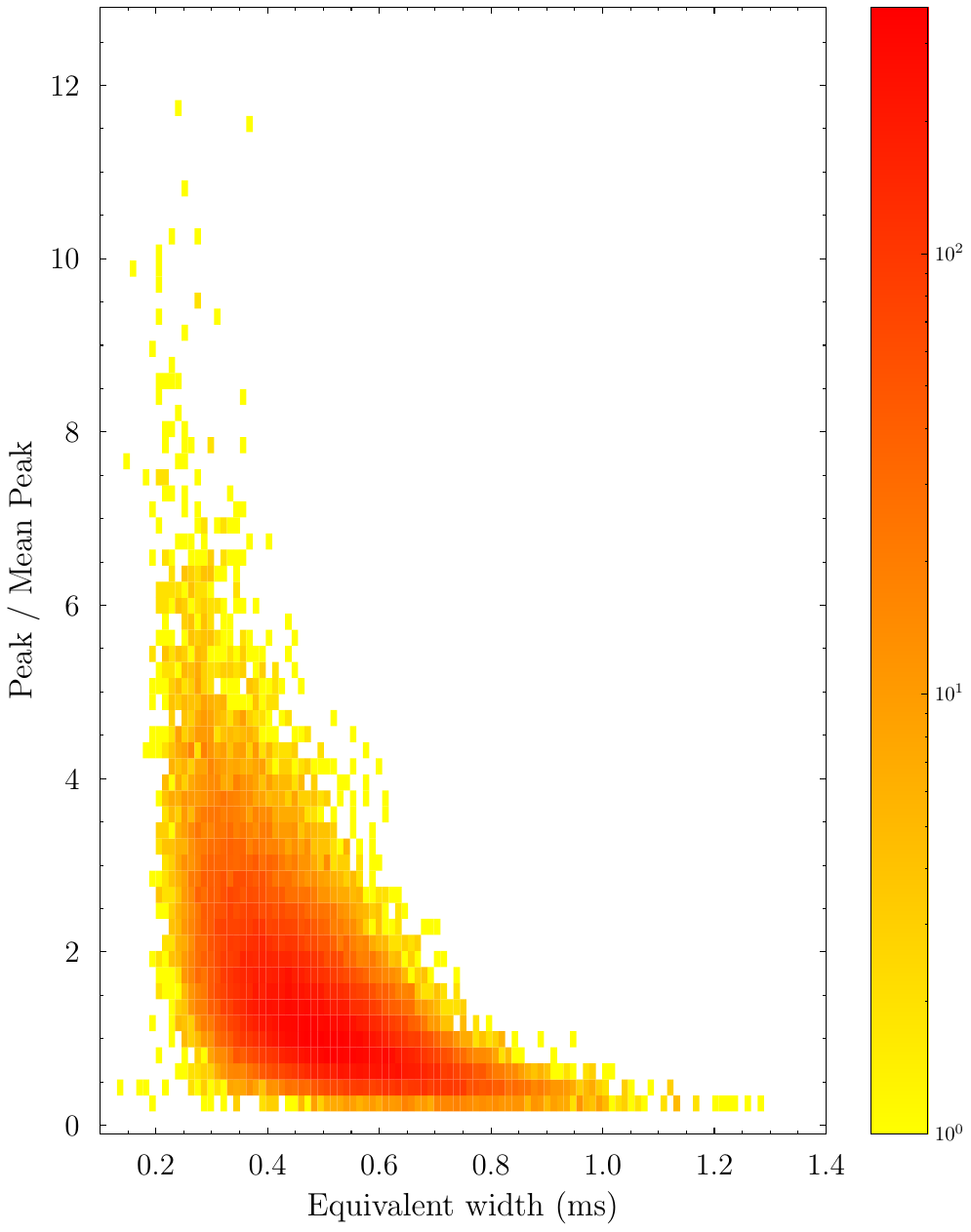}
    \caption{The number density distribution of pulses with respect to their relative peak flux densities and the equivalent width $W_{\rm eq}$.
    \label{fig:weq}}
\end{figure}

The energy distribution of the main-pulse region is shown in Fig.\,\ref{fig:energy}.
The energy distribution shows an obvious log-normal distribution.
The log-normal probability density function is defined as,
\begin{equation}\label{eq:lognormal}
P(E)=\frac{\langle E\rangle}{\sqrt{2 \pi} \sigma E} \exp \left[-\left(\ln \frac{E}{\langle E\rangle}-\mu\right)^{2} /\left(2 \sigma^{2}\right)\right]\,,
\end{equation}
where $\langle E\rangle$ is the average energy of the main pulse of PSR~J2222$-$0137; this corresponds to a SNR of $87$. 
Using the main-pulse energy distribution, we fit Eq.\,(\ref{eq:lognormal}) to the data to obtain the values of {$\mu=-0.1032\pm0.0021$} and {$\sigma=0.4108\pm0.0018$}. 
Performing a KS-test, we get a p-value, $p=0.81$, indicating that the log-normal distribution provides a very good description of the data. 
The fitted curve is the blue solid line in Fig.\,\ref{fig:energy}, where we see how closely the energy distribution of the main-pulse region conforms to a log-normal distribution. 
The off-pulse region follows a Gaussian distribution, corresponding to the noise's energy distribution.

The highest single pulse energy of the single pulse on MJD 59209 is about $6.9\,\langle E\rangle$. 
The definition of a giant pulse is a single pulse with energy above $10\,\langle E\rangle$ \citep{Cairns_2004ApJ}, so based on our results,
there is no giant pulse in the PSR J2222$-$0137 data taken on MJD 59209.
This is not unexpected since the pulse-energy distribution of giant pulses follows a power law.
The log-normal energy distribution of single pulses of PSR~J2222$-$0137 implies that the pulsar's radiation follows a stochastic growth theory prediction \citep{1992SoPh_Robinson,Cairns_2001}.

We compare the average pulse profiles selected from different ranges of SNR in Fig.\,\ref{fig:snrdiffer},
{and the definition of SNR in Fig.\,\ref{fig:snrdiffer} is Eq.\,({\ref{eq:snrdefine}}).}
The black solid line is the total intensity profile. 
We can notice that the average pulse profile built using the brightest pulses is narrower than the total average profile, and the phase of the leading peak is earlier than the phase of the leading component of the main peak of the pulse profile (the grey solid vertical line).
As the SNR of the single pulses decreases, the pulse profile resembles the total shape more and more, as we should expect, given that the weaker pulses are more numerous and contribute more energy to the pulse profile.
This phenomenon indicates that the high SNR single pulses may have an earlier arrival time than the low SNR single pulses for PSR~J2222$-$0137.

\begin{figure}
    \includegraphics[width=9cm]{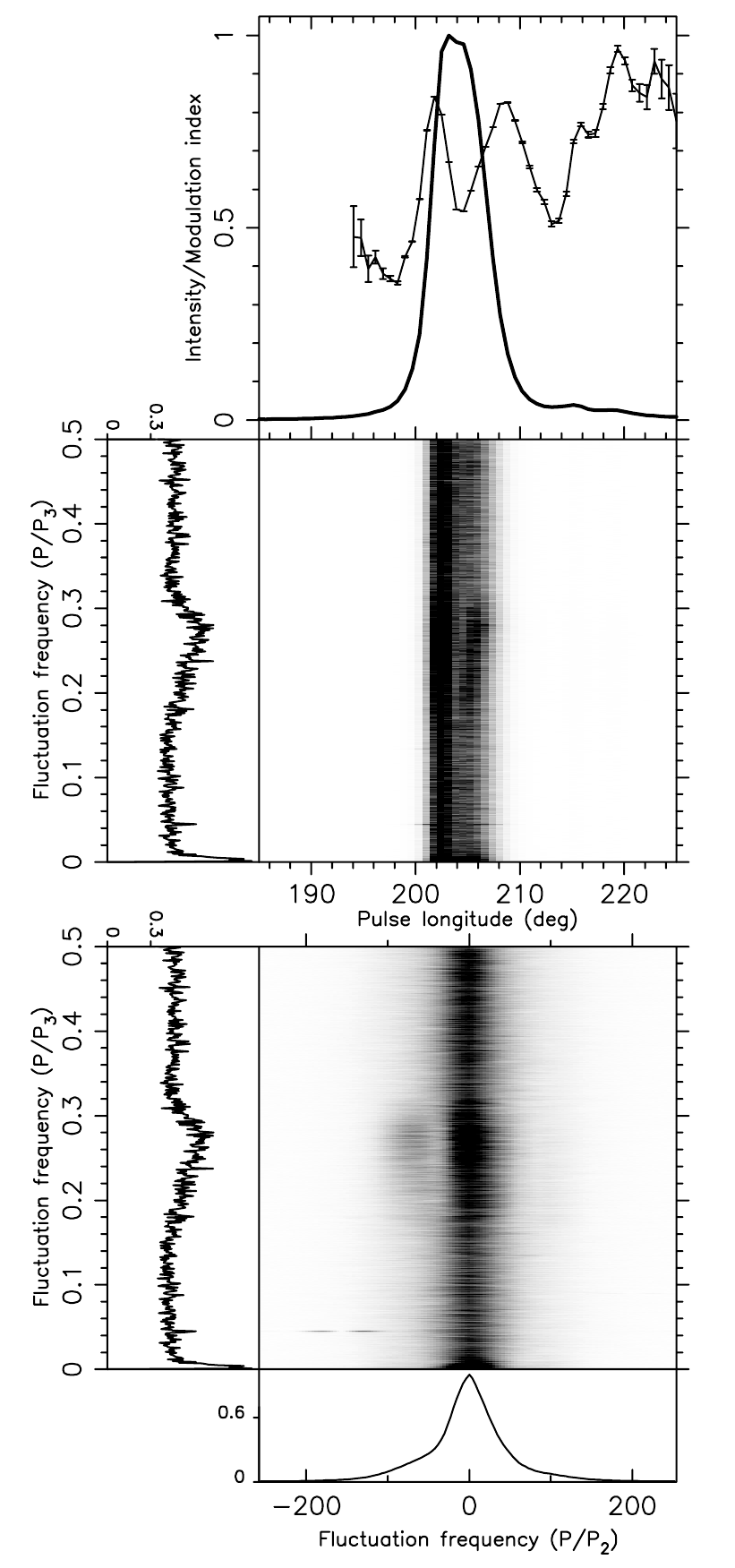}
    \caption{The top panel shows how the modulation index $m_{i}$ varies with the spin phase, the solid line is the integrated profile, and the points with error bars are $m_{i}$. The middle panel is the LRFS. 2DFS is below the LRFS, and the power in the 2DFS is vertically integrated and normalized with respect to the peak to obtain the bottom panel. Both the LRFS and the 2DFS are horizontally integrated and normalized with respect to the peak to obtain the left panels of the spectra.
    \label{fig:fourier}}
\end{figure}
\begin{figure*}
    \includegraphics[width=14cm]{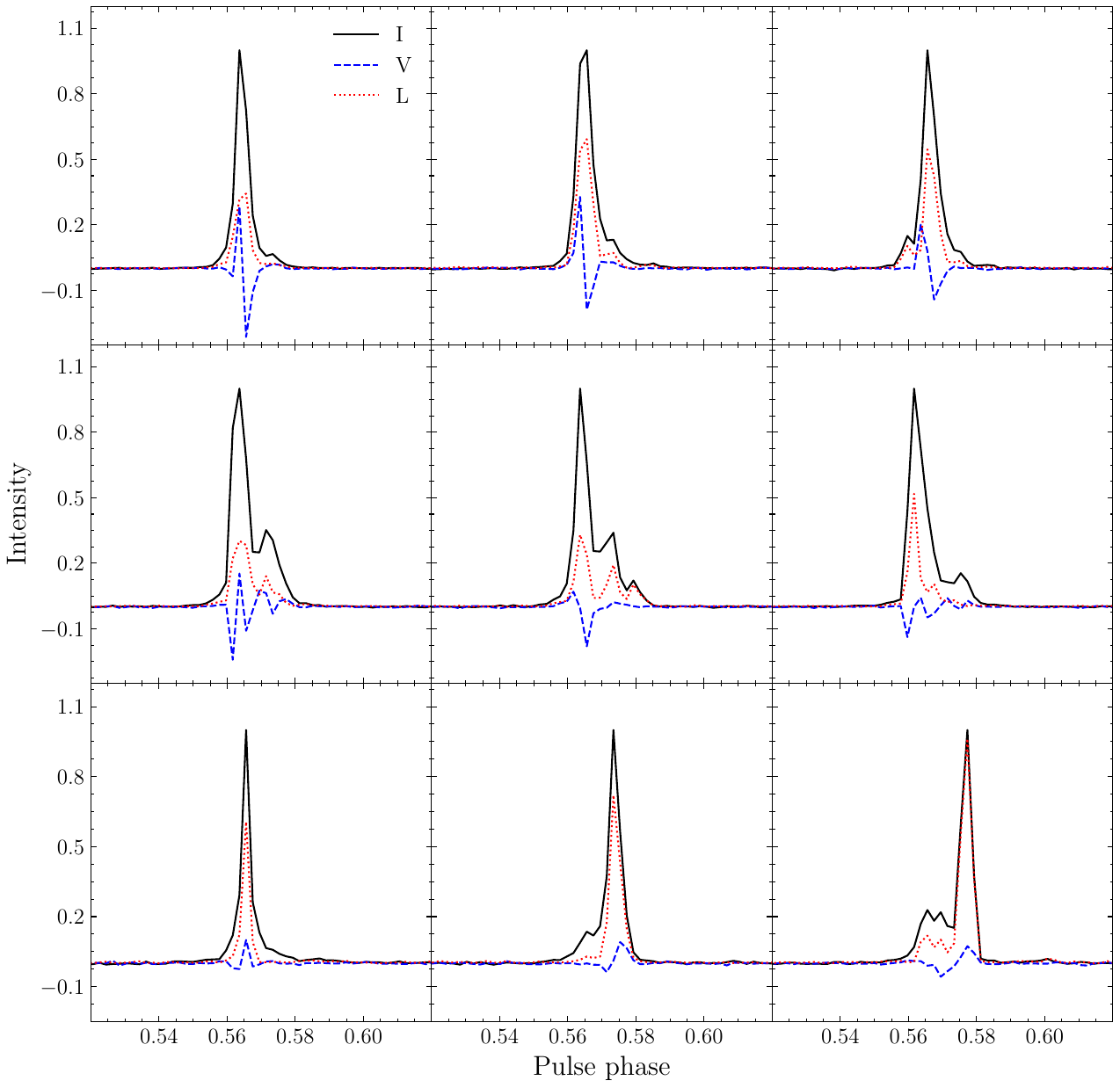}
    \caption{The nine high SNR single-pulse polarization profiles. The black solid line is the total intensity I, the red dotted line is linear polarization intensity L and the blue dashed line is circular polarization V. The time resolution in this figure is $64.06\,{\rm \mu s}$. \label{fig:ninepolar}}
\end{figure*}
We provide the number density distribution of pulses with respect to their relative peak flux densities and spin phases where the pulse peaks occur in Fig.\,\ref{fig:peakdis}.
Here, the pulses with higher peak flux densities concentrate in a narrow region corresponding to the leading peak of the total average profile.
The pulses with lower peak flux densities are distributed in a wider region.

We also plot the number density distribution of pulses in terms of their relative peak flux densities and the equivalent width $W_{\rm eq}$ in Fig.\,\ref{fig:weq}, and we define $W_{\rm eq}$ as the area under the pulse divided by the peak amplitude,
\begin{equation}\label{eq:weq}
    W_{\rm eq} = \frac{\sum I_{i}}{I_{p}}\times \Delta t,
\end{equation}
where {$I_{p}$ is the peak amplitude of single pulse and $\Delta{t}=64.06\,{\rm \mu s}$ is the time resolution.}
As can be seen, the $W_{\rm eq}$ of the main pulse of PSR~J2222$-$0137 concentrates in about $0.5\,{\rm ms}$.
The bright pulses have smaller $W_{\rm eq}$, $W_{\rm eq}<0.4\,{\rm ms}$, which is consistent with the result shown in Fig.\,\ref{fig:snrdiffer} that high SNR pulses have a narrower profile.
The weak pulses have a wider $W_{\rm eq}$ distribution.

{In general, the position of each individual single pulse is random.
If pulsars have sub-pulse drifting, the pulse-to-pulse variation is known to be regular.}
In order to study the properties of variation of the main pulse, we have calculated the longitude-resolved modulation index that measures the normalized RMS intensity across pulse phase $i$ and it is defined as \citep{shannon_2012ApJ},
\begin{equation}\label{eq:modulation}
    m_{i,\,\rm {on}}=\frac{\sqrt{\sigma_{i,\,\rm {on}}^{2}-\sigma_{\rm off}^{2}}}{I_{i,\,\rm on}},
\end{equation}
where $I_{i,\,\rm on}$ and $\sigma_{i,\,\rm {on}}$ are the mean and RMS intensity at pulse phase $i$ of on-pulse region.
We plot the modulation index $m_{i}$ as a function of the pulse phase of PSR~J2222$-$0137 in the top figure of Fig.\,\ref{fig:fourier}.
Except for the edge region of the integrated profile of the main-pulse region, the modulation indices are lower than 1.
The values of $m_{i}$ at the edges of the integrated profile indicate a significant change in the strength of pulses, commonly seen in normal pulsars.
The large value of $m_{i}$ near the leading peak edge indicates the existence of a tail of bright pulses, which is consistent with the conclusion of Fig.\,\ref{fig:snrdiffer}.
The modulation index $m_{i}$ has an asymmetric distribution, meaning that the leading and trailing edges of the integrated profile have a different intensity variation, which is similar to PSR J1713+0747 \citep{2016MNRAS_Liu}.

{We use PSRSALSA to obtain the power spectra of the pulse intensity as a function of pulse number and rotational phase, which can reflect the periodicities in the radio emission. This is known as the Longitude Resolved Fluctuation Spectrum (LRFS), which can identify the fluctuation spectrum of constant pulse longitude in the pulse stacks \citep{1970Natur_backer}. 
The LRFS uses a Discrete Fourier Transform (DFT) to reveal the presence of periodicities for each pulse longitude bin.
The middle panel in Fig.\,\ref{fig:fourier} shows the LRFS result,
where the horizontal axis is $P/P_{3}$, and $P$ is the pulsar spin period, and $P_{3}$ is the sub-pulse pattern repetition rate, namely the period of sub-pulse drifting.}
There are two maxima values, one at 0.28 cycles-per-period (cpp) and one near 0 cpp.
{The LRFS is a one-dimensional DFT, and it can not distinguish whether periodic changes of a constant phase bin are due to intensity modulation or sub-pulse drifting.

To distinguish between these possibilities, we must introduce the Two-Dimensional Fluctuation Spectrum (2DFS), obtained by performing DFTs along lines with different slopes in the pulse stack \citep{2002A&A_Edwards}.
The 2DFS does not only give information about the periodic intensity variation along the pulse stack but also provides information on the horizontal separation of the drifting in pulse longitude, namely $P_{2}$.}
We calculate the 2DFS and plot it in the bottom figure of Fig.\,\ref{fig:fourier}.
We note an offset from the vertical axis of zero at 0.28 cpp ({corresponding} {$P_{3}\simeq3.5\,P$}) in the bottom panel of Fig.\,\ref{fig:fourier}, corresponding to a possible systematic drifting of emission power in the pulse phase.
The horizontal separation of the drifting in pulse longitude $P_{2}\sim -5.8^{\circ}$.
The values of $P_{2}$ and $P_{3}$ are not evident. So there is a weak systematic drifting of emission power in the pulse phase.

\begin{figure}
    \includegraphics[width= \columnwidth]{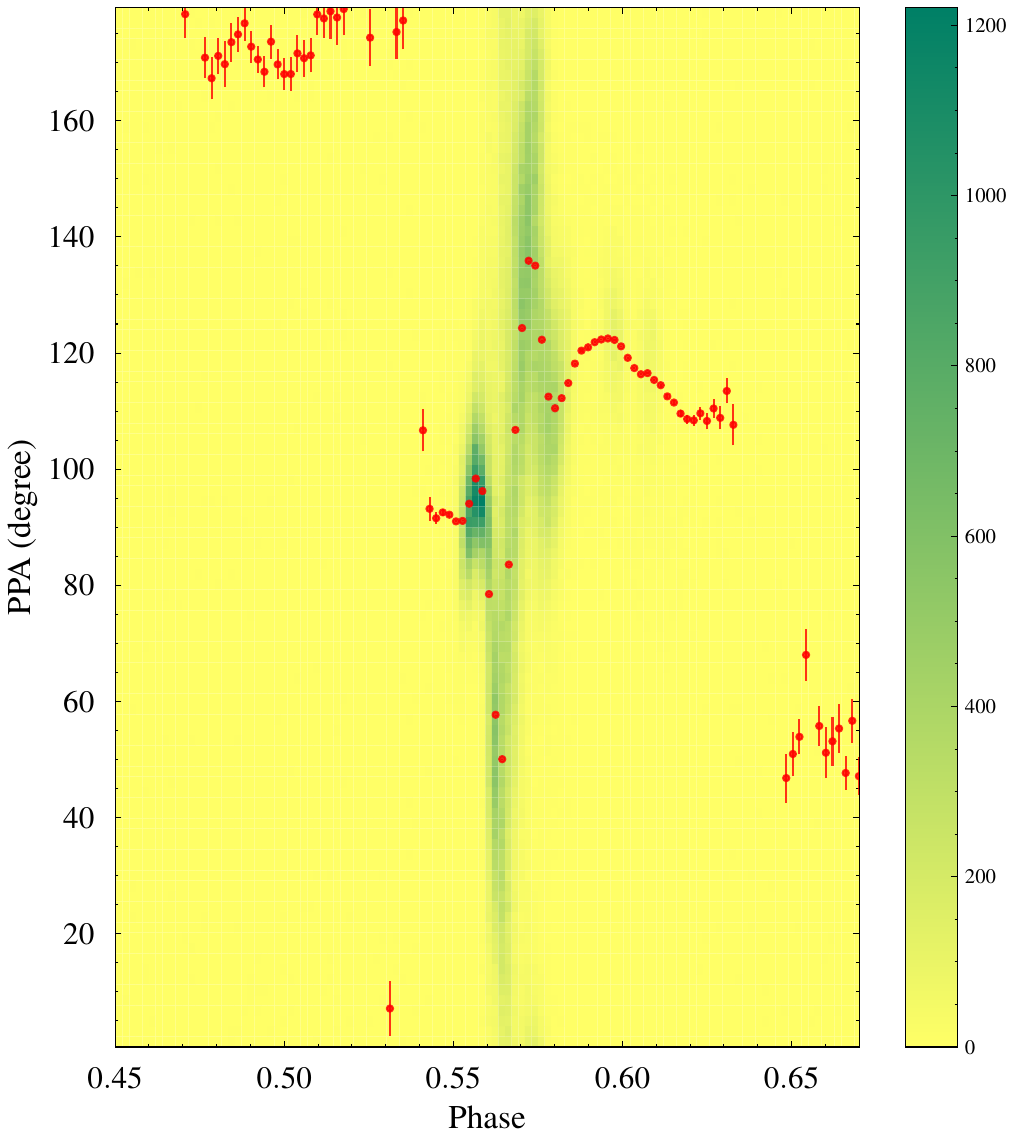}
    \caption{The longitude-resolved probability density distribution of {PPA} values from single pulses of PSR~J2222$-$0137. The values are selected if the corresponding linear polarization ${\rm SNR}>15$ and the error of {PPA} is less than $10^{\circ}$.
    The red dots with the error bar are the {PPA} values from the integrated profile. We only plot the spin phase between $0.52-0.62$, corresponding to the region of rapid changes in {PPA} and V for the integrated polarization profile.\label{fig:pa_angle}}
   
\end{figure}

\subsection{Single pulse polarization}
\label{sec:res:polar}

The MJD 59209 observation can help us to investigate the polarization properties of single pulses from MSPs.
We use the Stokes parameters I, Q, U, and V to describe the polarization of single pulses.
I is the total intensity, V is the intensity of circular polarization, and ${\rm L}=\sqrt{{\rm Q^{2}}+{\rm U^{2}}}$ define the linear polarization.
We plot nine single-pulse polarization profiles in Fig.\,\ref{fig:ninepolar}.
The high SNR of these single pulses provides transparent information on circular and linear polarization. 
{In Fig.\,\ref{fig:ninepolar}, the top and middle panels show single pulses with a rapid sign change of circular polarization.}
We note that for some single pulses {of PSR J2222$-$0137}, the timescale for the change in circular polarization from a negative to a positive peak is about $0.15$ ms.
Such rapid sign change of circular polarization may originate from an intrinsic mechanism, e.g., the LOS sweep across the central trajectory of charged bunches \citep{Wang22}. 
It could also be caused by propagation effects {\citep{Sazonov_1969SvA,Melrose_1997PhRvE}}.
{In Fig.\,\ref{fig:ninepolar}, the bottom panel shows single pulses with high linear polarization.}

As mentioned above, \cite{Guo_2021} have first analysed the integrated polarization profile of PSR J2222-0137 with the data used in this paper.
In Fig.\,3 of that work, the bottom panel shows values of the {PPA} of the linear emission as a function of the longitude. The authors used a RVM to fit the black {PPA} values,
and the gray {PPA} points, which correspond to the rapid changes in {PPA}, are not fitted.
The rapid change region of {PPA} corresponds to the sudden drops in linear polarization intensity and the very rapid variation in {circular polarization intensity}, and the authors suggested that this may imply the existence of OPMs. 

To reveal if there are OPMs in the grey region, we use the same data to calculate the {PPA} distribution of single pulses of PSR~J2222$-$0137.
Fig.\,\ref{fig:pa_angle} shows the resulting {PPA} distribution with the spin phase corresponding to the grey region in Fig.3 of \citet{Guo_2021}.
The red dots are the {PPA} values of the integrated profile. We note that the {PPA} distribution of single pulses shows the same trend as the {PPA} of the integrated profile.
There is no evidence for discontinuities of approximately $90^{\circ}$ of the single pulse {PPA} values between spin phase $0.52-0.62$ at $64.06\,{\rm \mu s}$ time resolution.
From these data, it is not obvious that OPMs for PSR~J2222$-$0137 exist in this spin phase region. However, a steep transition between modes could still exist at higher time resolutions.

\subsection{Single pulse jitter}
\label{sec:res:jitter}
{For a same integration time scale, MSPs have more stable integrated pulse profiles than normal pulsars, enabling precision TOAs. This greater stability may be simply due to the much larger number of pulses added for a same time (see chapter 8 in \citealt{handbook_2005}). For MSPs' single pulses,}
the shape and arrival phase of pulse to pulse could change dramatically, called the ``pulse jitter''.
The existence of jitters can cause TOA fluctuations even when derived from integrated profiles, contributing to an additional TOA measurement error.
We can express the relation as \citep{Liu_2011},
\begin{align}
    \sigma_{\rm obs}^{2}=\sigma_{\rm rn}^{2}+\sigma_{\rm J}^{2}+\sigma_{\rm sc}^{2}\,,
\end{align}
where $\sigma_{\rm obs}$ is the actual measurement uncertainty of the TOA, $\sigma_{\rm rn}$ is the uncertainty induced by radiometer noise, $\sigma_{\rm J}$ is the uncertainty induced by pulse jitter and $\sigma_{\rm sc}$ is induced by instability of short-term diffractive scintillation.
For PSR~J2222$-$0137, the contribution of $\sigma_{\rm sc}$ can be ignored\footnote{The scintillation frequency is {$\Delta \nu=30.7\pm6.5\,{\rm MHz}$} and the scintillation time scale is {$\Delta t=11.1\pm2.3\,{\rm min}$}. We use the Eq.(4) in \citet{shannon_2012ApJ}, and for 30-s integration time, we get the rms time error contributed by diffractive scintillation, $\Delta t_{\rm DISS}\sim 3\,{\rm ns}$.}.
The Eqs.\,(4) and (5) in \citet{Liu_2011} provide the formalism of $\sigma_{\rm rn}$ and $\sigma_{\rm J}$, and they are all proportional to $1/\sqrt{N}$, namely, $\sigma_{\rm obs}^{2}(N)=\sigma_{\rm rn}^{2}(N)+\sigma_{\rm J}^{2}(N)$, where $N$ is the integrated number of pulses.
\begin{figure}
    \includegraphics[width= \columnwidth]{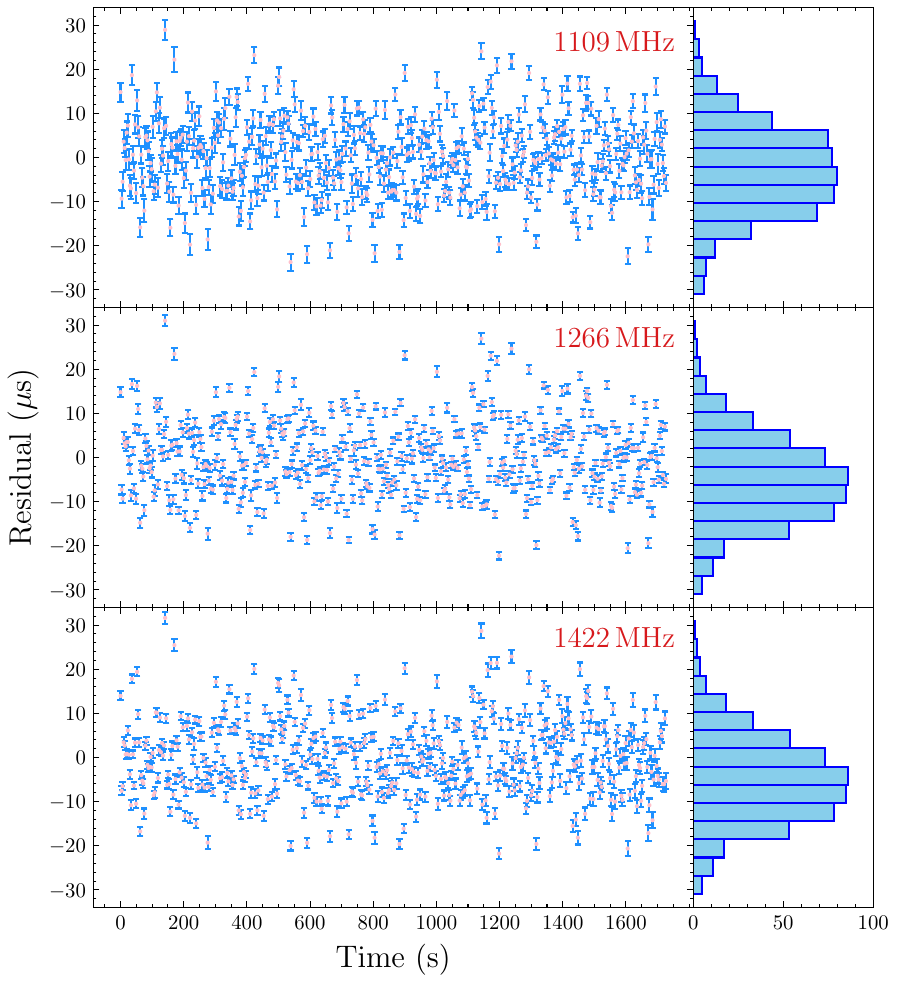}
    \caption{The left panels are residual TOAs for observations obtained simultaneously in three sub-bands of 31.25 MHz. The centre frequencies of the bands are 1109 MHz for the top panel, 1266 MHz for the middle panel, and 1422 MHz for the bottom panel. The right panels are histograms of the time residuals of the TOAs.
    \label{fig:fbandres}}
\end{figure}

\begin{figure}
    \includegraphics[width= \columnwidth]{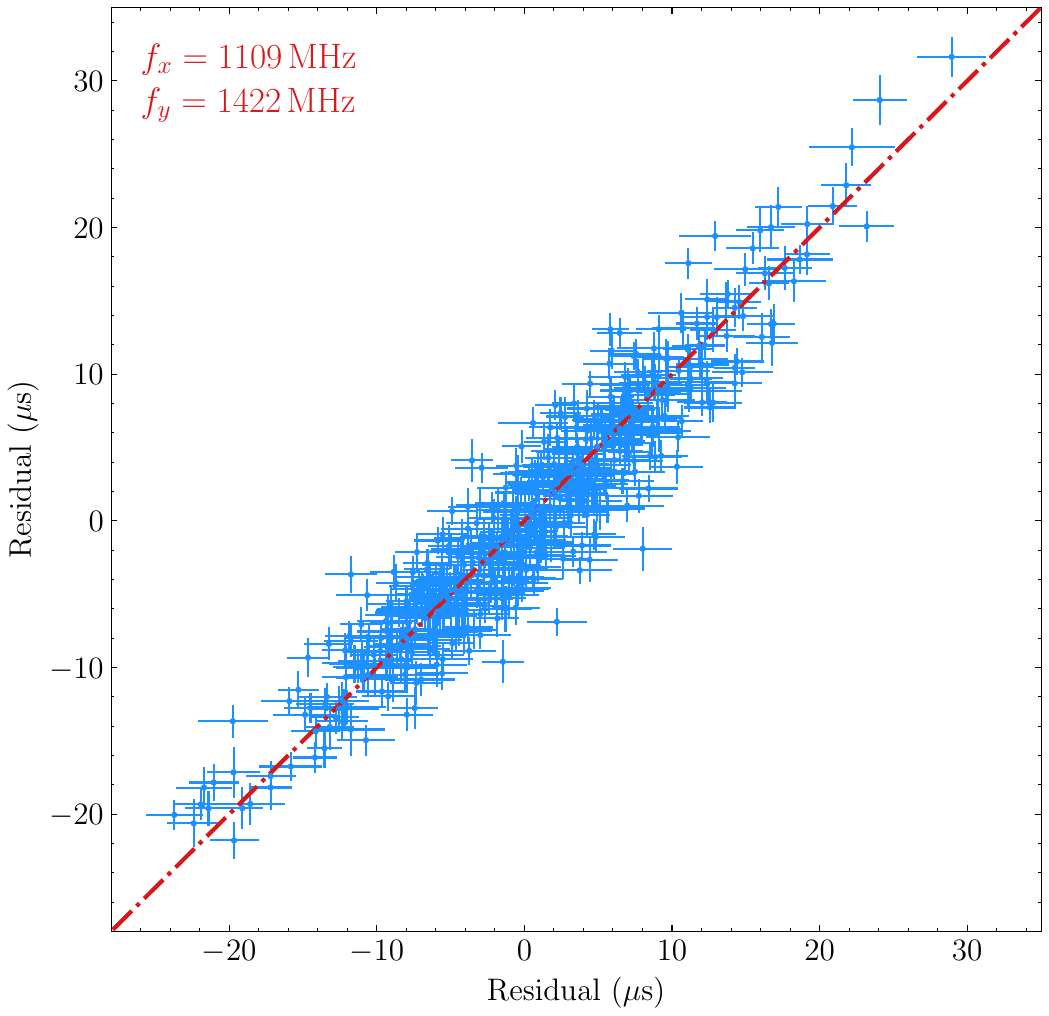}
    \caption{The correlation of the TOA residuals between the frequencies 1109 MHz and 1422 MHz. The error bars represent the 1-$\sigma$ template fitting errors. The red dashed line corresponds to the point where the residuals for the two frequency bands are equal.
    \label{fig:fbandcor}}
\end{figure}

\citet{shannon_2012ApJ} plotted the correlation of TOA timing residuals between different frequency sub-bands, showing high correlations. 
They suggested that such high correlations between different bands could not originate from radiometer noise and demonstrated that they should result from pulse-to-pulse jitter.
They calculated CCFs for the TOA residuals between the different sub-bands, and the value at zero lag reflected $\sigma_{\rm J}$, namely CCF(0)$=\sigma_{\rm J}^{2}$.
Similar to their work, we perform a correlation analysis for residuals at different frequencies.
We divide the $500$-MHz bandwidth into 16 sub-bands of width $\Delta f  = 31.25\,{\rm MHz}$.
We integrate 100 single pulses to form one TOA and use {\sc TEMPO} to get timing residuals of 16 frequency bands. 
Fig.\,\ref{fig:fbandres} illustrates the residuals in three different frequency bands.
{The errors of TOAs are much smaller than the residual scatter, so the radiometer noise makes a small contribution to $\sigma_{\rm obs}$.}
Fig.\,\ref{fig:fbandcor} shows
the correlation of the residuals between two frequency channels ($f_{x}=1109\,{\rm MHz},\, f_{y}=1422\,{\rm MHz}$).
We can observe a strong correlation between the residuals of the two frequency sub-bands. This indicates that the timing RMS of PSR~J2222$-$0137 is primarily influenced by jitter noise.

\begin{figure}
    \includegraphics[width= \columnwidth]{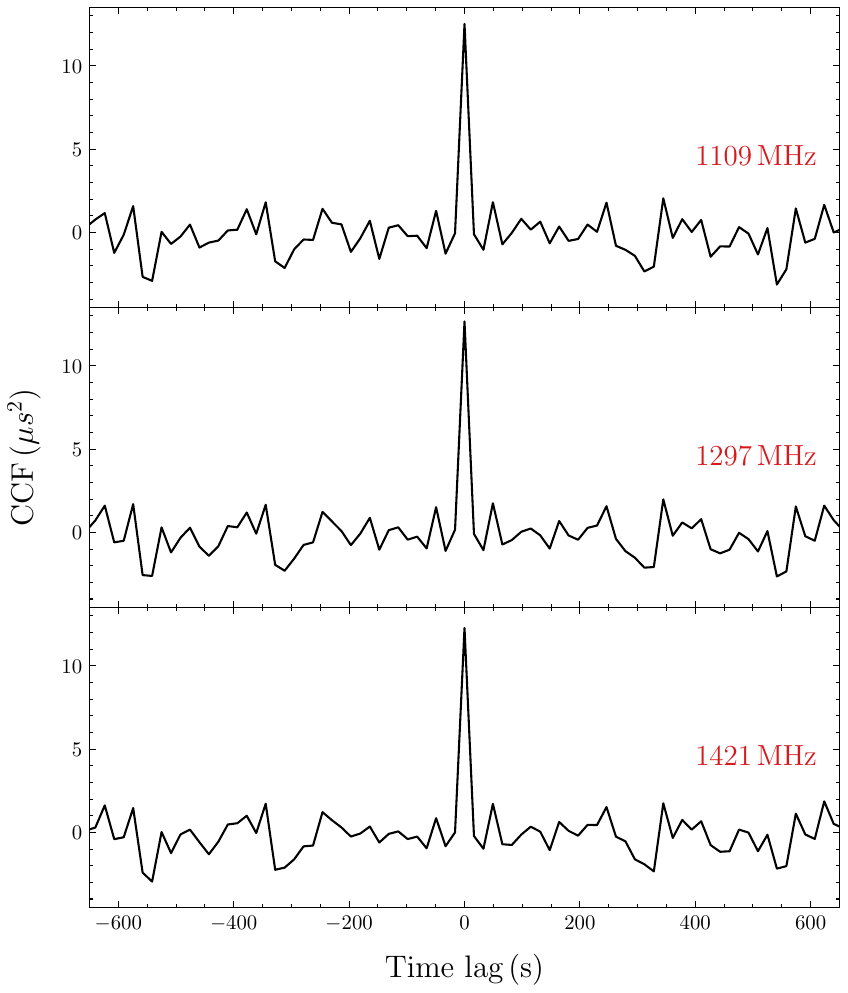}
    \caption{The cross-correlation function of 500 single-pulse integrations between time series for the center frequency 1266 MHz and three other frequency bands. \label{fig:fbandccf}}
\end{figure}

\begin{figure}
    \includegraphics[width= \columnwidth]{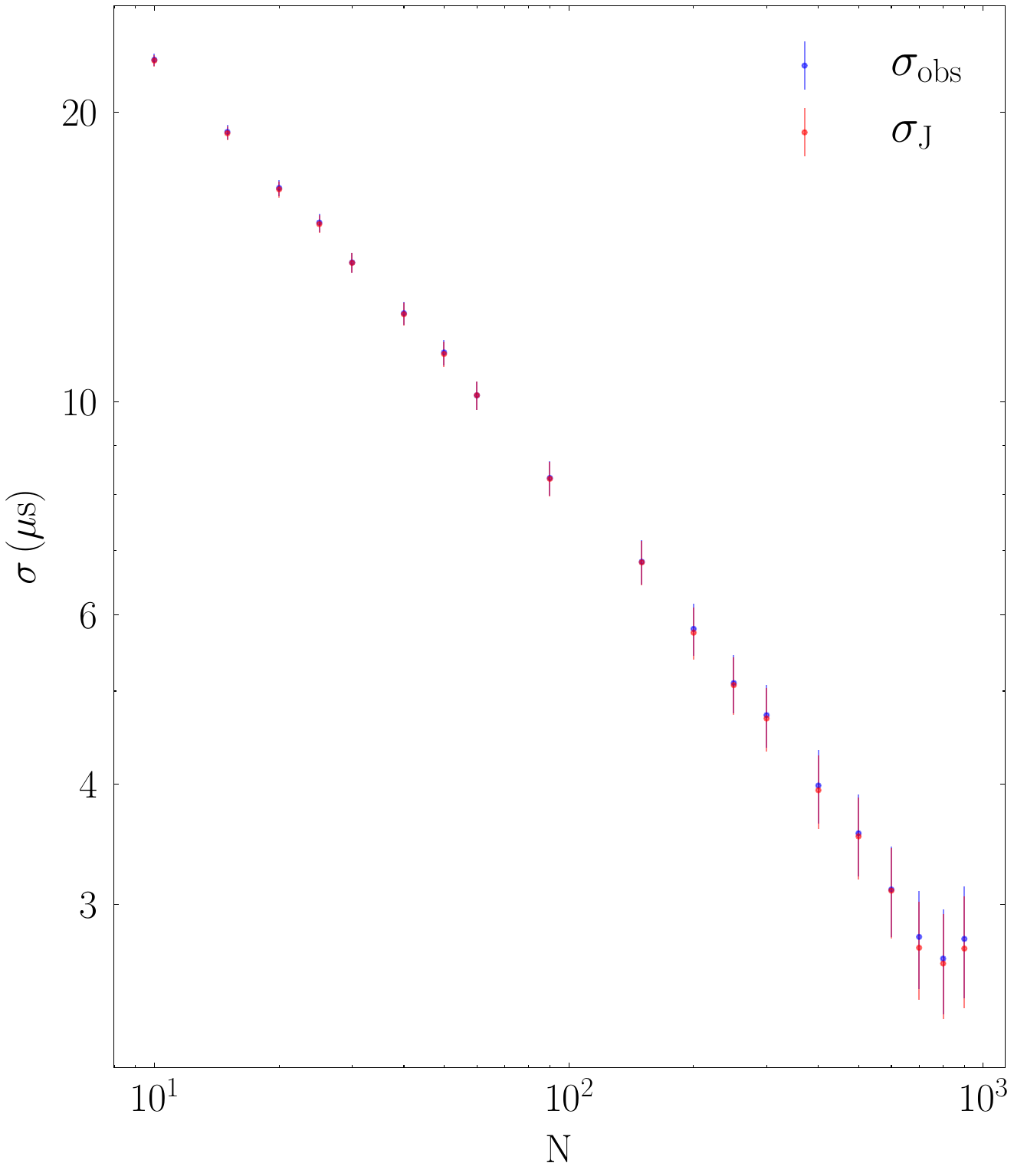}
    \caption{The $\sigma_{J}$ (red dots) calculated by CCF(0) {from different frequency bands ($\Delta f  = 31.25\,{\rm MHz}$). We take an average from the values of CCF(0) of different frequency bands for every fixed number of pulses averaged N.}  $\sigma_{\rm obs}$ (blue dots) variation with respect to the number of pulses averaged N. \label{fig:plotjitter_fand}}
\end{figure}

We calculate the CCF(0) between different frequency bands with 500 single-pulse integration (integration time $=16.4\,{\rm s}$), and the results are shown in Fig.\,\ref{fig:fbandccf}. 
CCF(0)=$12.47\,\mu s^{2}$, namely $\sigma_{J}(500)\approx3.53\,{\rm \mu s}$, corresponding to jitter of single pulses of $\sqrt{500}\times 3.53\,{\rm \mu s}\approx 78.93\,\mu {\rm s}$.
We use this method to calculate $\sigma_{\rm J}$ for different numbers $N$ of averaged pulses, and the results are shown in Fig.\,\ref{fig:plotjitter_fand}.
It confirms that jitter noise is the main contributor to $\sigma_{\rm obs}$.

\begin{figure}
    \includegraphics[width= \columnwidth]{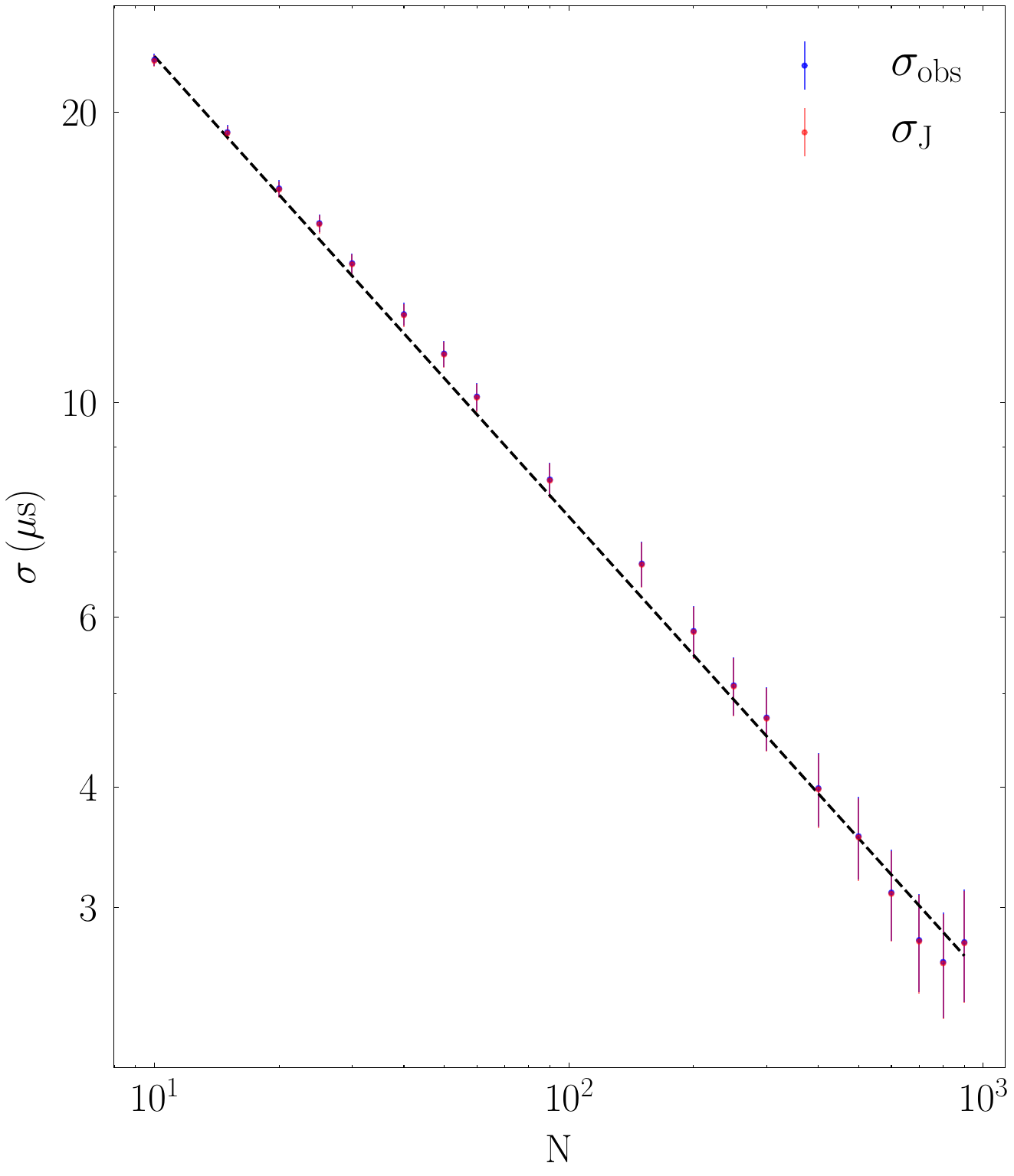}
    \caption{The $\sigma_{J}$ (red dots) calculated by Eq.\,(\ref{eq:calcu_jitter}) and $\sigma_{\rm obs}$ (blue dots) variation with respect to the number of pulses averaged N.
    {The variation of $\sigma_{J}$ is same with $\sigma_{J}$ calculated by CCF(0) of different frequency bands.}
    The black dashed line is a fit to $\sigma_{\rm J}$ with $\sigma_{\rm J}(N)=\alpha N^{\beta}$, and we get 
    $\alpha=68.52\pm1.48\,{\rm\mu s}$ and $\beta=-0.48\pm0.01$ which are similar with the expected scaling $\sigma_{\rm J} \propto 1/\sqrt{N}$.
    \label{fig:plotjitter_rn}}
\end{figure}

On the other hand, $\sigma_{\rm obs}(N)$ is related to $\sigma_{\rm rn}(N)$ and $\sigma_{\rm J}(N)$, so we can calculate $\sigma_{\rm J}(N)$ by,
\begin{align}
    \sigma_{\rm J}(N)=\sqrt{\sigma_{\rm obs}^{2}(N)-\sigma_{\rm rn}^{2}(N)}\,.\label{eq:calcu_jitter}
\end{align}
We calculate $\sigma_{\rm rn}^{2}(N)$ for different pulse numbers $N$, and use Eq.\,(\ref{eq:calcu_jitter}) to get $\sigma_{\rm J}(N)$.
We plot results in Fig.\,\ref{fig:plotjitter_rn}, and note that the variation in $\sigma_{\rm J}$ with $N$ is consistent with Fig.\,\ref{fig:plotjitter_fand}.
Hence, two methods produce consistent results for $\sigma_{\rm J}(N)$.
Theoretically, $\sigma_{\rm J}(N)$ is proportional to $1/\sqrt{N}$. 
We use $\sigma_{\rm J}(N)=\alpha N^{\beta}$ to fit the variation with $N$ {in Fig.\,\ref{fig:plotjitter_rn}}, and get $\alpha=68.52\pm1.48\,\mu{\rm s}$ and $\beta=-0.48\pm0.01$, confirming that is satisfies $\sigma_{\rm J} \propto 1/\sqrt{N}$.
With values of $\alpha$ and $\beta$, for such high SNR, we can predict jitter noise at $1\,$hr,
$\sigma_{\rm J}(1\,{\rm hr})=270\pm 9\,{\rm ns}$.
So, for high SNR data of PSR J2222$-$0137 from FAST, 
the TOAs’ noise is dominated by jitter noise.

Despite this, FAST timing is still more precise than the observations taken by other telescopes.
For 15-min integrations, we can get $\sigma_{\rm J}(15\,{\rm min})=519 \pm 20\,{\rm ns}$, the radiometer noise $\sigma_{\rm rn}(15\,{\rm min})$ is {$\sim140\,{\rm ns}$ (mean value)}. Thus, for 15-minute observations, the TOA precision is jitter-dominated. Nevertheless, it is still substantially better than what has been achieved to date:
Table 2 of \citet{Guo_2021} provides the weighted residual RMS $\sigma_{\rm obs}=2.27\,\mu {\rm s}$ for 900 s TOAs from the Effelsberg telescope at a 1.4 GHz frequency band.
Hence, despite the fact that FAST observations of PSR J2222$-$0137 are affected by jitter noise, they will still represent a significant improvement in the timing precision for this pulsar.


\section{Conclusion and discussion}
\label{sec:con}
In our work, we report the study of $52692$ single pulses from a high-SNR 30-minute observation of the recycled pulsar PSR~J2222$-$0137 and analyse their statistical properties.
The energy distribution of the main-pulse region shows a log-normal distribution, which means that this system has no detectable nulling, mode changes, or giant pulses.
We get the LRFS and 2DFS, and the results suggest a weak sub-pulse drifting in the main pulse of PSR~J2222$-$0137, with $P_{2}\sim-5.8^{\circ}$ and {$P_{3}\sim3.5\,P$}.
We analyse the polarization of single pulses of the main-pulse region of PSR~J2222$-$0137 and find no evidence for OPMs in the spin phase $0.52-0.62$ within $64\,{\rm \mu s}$ time resolution.
There could be a more complicated emission or propagation process leading to the sudden drops in linear polarization intensity and the very rapid variation of {circular polarization intensity in the $0.52-0.62$ spin phase, or the existence of OMPs visible at higher resolution.

We calculate the jitter noise of PSR~J2222$-$0137 with different numbers of integrated pulses.
The cross-correlation between different frequency bands and using Eq.\,(\ref{eq:calcu_jitter}) both provide a similar variation relation with $N$, and the relation is consistent with $\sigma_{J}\propto1/\sqrt{N}$.
The FAST timing data from PSR~J2222$-$0137 is clearly jitter-dominated; in other words, jitter noise is now the main factor limiting the timing precision, especially for the observations where scintillation significantly brightens the pulsar.
When this happens, increases in the sensitivity of the radio telescope can not improve the timing precision.
We need to use longer integration times to enhance the timing precision. 
At 1.25 GHz, we estimate that the jitter noise contribution is $\approx270\,{\rm ns}$ for 1-h integration. Despite the fact that jitter disproportionately affects the FAST data on PSR~J2222$-$0137, especially for the brightest observations, the FAST data still significantly improve the overall timing precision.

\citet{batrakov2023new} use the TOA uncertainties of PSR~J2222$-$0137 from FAST to simulate a 10-yr timing campaign of PSR~J2222$-$0137. They use a 15-min integration time to reduce the jitter effect because it is jitter-dominated. They predict that the FAST timing of PSR~J2222$-$0137 will result in much-improved measurements of the masses and geometric parameters and vastly improve the constraints on the scalar-tensor theories obtained with this system.
So, by reasonably arranging timing campaigns of PSR~J2222$-$0137 with FAST, which takes the information on the timing precision into account, these timing data can play an important role in improving the ability to test gravity theories with this system.


\section*{Acknowledgements}

{We thank the anonymous referee for constructive comments that improved the
work. We thank Kuo Liu,} Jinchen Jiang, Heng Xu, Dejiang Zhou, Weiyang Wang, Feifei Kou, Yi Feng and Shi Dai for valuable discussions and suggestions. Xueli Miao is supported by the Cultivation Project for FAST Scientific Payoff and Research Achievement of CAMS-CAS. This work was supported by the National SKA Program of China (2020SKA0120200, 2020SKA0120300), 
the National Natural Science Foundation of China (12203072, 12041303, 12203070), and the CAS-MPG LEGACY project.
Jumei Yao is supported by the Natural Science Foundation of Xinjiang Uygur Autonomous Region (grant No. 2022D01D85) and the Major Science and Technology Program of Xinjiang Uygur Autonomous Region (grant No. 2022A03013-2).

\section*{Data Availability}

Origin raw data are published by the FAST data center and can be accessed through them. For other intermediate-process data, please contact the author (xlmiao@bao.ac.cn).



\bibliographystyle{mnras}
\bibliography{ref} 








\bsp	
\label{lastpage}
\end{document}